\documentclass[]{emulateapj}
\slugcomment{Revised manuscript submitted to ApJ}

\shorttitle{CORONAL HARD X-RAY SOURCES} \shortauthors{CHEN \& PETROSIAN}

\begin{document}

\title{IMPULSIVE PHASE CORONAL HARD X-RAY SOURCES IN AN X3.9 CLASS SOLAR FLARE}
\author{Qingrong Chen$^{1}$ and Vah\'e Petrosian$^{1,\ 2}$}
\affil{$^{1}$
Department of Physics and KIPAC, Stanford University, Stanford, CA 94305, USA; 
qrchen@gmail.com, vahep@stanford.edu}
\affil{$^{2}$
Department of Applied Physics, Stanford University, Stanford, CA 94305, USA.}

\begin{abstract}
We present analysis of a pair of 
unusually energetic coronal hard X-ray (HXR) sources detected by 
the {\it Reuven Ramaty High Energy Solar Spectroscopic Imager} ({\it RHESSI})
during the impulsive phase of an X3.9 class solar flare on 2003 November 3,
which simultaneously shows two intense footpoint (FP) sources.
A distinct loop top (LT) coronal source is detected up to $\sim$150 keV 
and a second (upper) coronal source up to $\sim$80 keV. 
These photon energies, 
which were not fully investigated in earlier analysis of this flare,
are much higher than commonly observed in coronal sources 
and pose grave modeling challenges.
The LT source in general appears higher in altitude 
with increasing energy and exhibits a more limited motion 
compared to the expansion of the thermal loop. 
The high energy LT source shows an impulsive time profile and
its nonthermal power law spectrum exhibits soft-hard-soft evolution 
during the impulsive phase, similar to the FP sources.
The upper coronal source exhibits an opposite spatial gradient 
and a similar spectral slope compared to the LT source. 
These properties are consistent with the model of stochastic acceleration 
of electrons by plasma waves or turbulence.
However, the LT and FP spectral index difference
(varying from $\sim$0--1) 
is much smaller than commonly measured and than that expected from a simple 
stochastic acceleration model. 
Additional confinement or trapping mechanisms of high energy electrons 
in the corona are required.
Comprehensive modeling including both kinetic effects and the
macroscopic flare structure may shed light on this behavior. 
These results highlight the importance of imaging spectroscopic observations
of the LT and FP sources up to high energies 
in understanding electron acceleration in solar flares.
Finally, we show that 
the electrons producing the upper coronal HXR source 
may very likely be responsible for the type III radio bursts 
at the decimetric/metric wavelength observed 
during the impulsive phase of this flare.
\end{abstract}


\section{Introduction}

It is generally believed that 
during the impulsive phase of solar flares, electrons are often accelerated 
to hundreds of keV (and sometimes to relativistic energies) as a result of 
energy release by magnetic reconnection.
However, the exact mechanisms of particle acceleration are still 
under much debate \citep[e.g.][]{Miller97, Krucker08a, Zharkova11}.
These nonthermal electrons are most directly connected to
the HXR emission they produce
through the well-known bremsstrahlung process \citep[e.g.][]{Lin74}.
Direct detection of HXR sources in the corona 
is thus of paramount importance 
to study the energy release and particle acceleration processes. 
The flare accelerated electrons attached to open field lines
will escape from the Sun and produce type III radio bursts 
and may be detected {\it in situ} by space instruments.

HXR imaging observations have shown that for most solar flares,
the majority of impulsive phase nonthermal emission 
comes from the conjugate FP regions
of a closed loop or arcade structure 
\citep[e.g.][]{Hoyng81, Sakao94, Saint-Hilaire08}.
This has been interpreted in terms of the collisional thick target model
\citep{Brown71, Syrovat-Skii72, Hudson72, Petrosian73},
in which nonthermal electrons move downward 
from the corona to the chromosphere 
and radiate most of the HXR emission at the dense FP regions.
This simple model with a beam of electrons injected into a coronal loop
does not predict a distinct HXR source from the tenuous corona 
except at very low energies, 
say below $\sim$20 keV \citep[e.g.][]{Leach83, Brown02}.
However, within the past two decades 
distinct coronal HXR sources have been found around the top of 
a thermal soft X-ray loop in addition to two FP sources,
with the first definitive observation
made by the {\it Yohkoh}/Hard X-ray Telescope (HXT) 
up to 33--53 keV \citep{Masuda94, Masuda95}. 
Observations of distinct coronal sources have motivated several models 
in terms of particle acceleration and/or transport effects 
\citep[see review by][]{Fletcher99},
assuming different properties of a coronal loop or the accelerated electrons,
such as a high loop density \citep{Wheatland95, Holman96},
magnetic field convergence \citep[][see also \citealt{Leach84}]{Fletcher98},
and plasma turbulence \citep{Petrosian99}.

Investigations of the {\it Yohkoh} and {\it RHESSI} flares 
have shown that coronal HXR emission is a common feature of all flares
\citep[e.g.][]{Petrosian02, Jiang06, Krucker08c}.
Analysis of the flares with simultaneously detected LT and FP sources 
\citep[e.g.][]{Petrosian02, BB06, Shao09} indicate that in general 
the LT spectra are much softer than the FP spectra and can be fitted 
by a relatively steep power law (sometimes plus a lower energy thermal component).
This fact, jointly with the finite dynamic range 
($\sim$10:1 for {\it Yohkoh}/HXT and {\it RHESSI}),
may explain why the LT sources are difficult to detect 
above $\sim$30 keV when the stronger FP sources are in the field of view.
Thus in-depth studies of the coronal LT emission largely come from 
the partially occulted solar flares, 
in which the more intense FP sources are blocked by the solar limb.
In these flares, the HXR spectra are found to be generally much softer than 
those non-occulted flares \citep{Krucker08c, Tomczak09}.
On the other hand, some peculiar conditions may yield much stronger
coronal bremsstrahlung sources than commonly seen.
For example, unusually dense loops can prevent nonthermal electrons 
reaching the FP regions so that the HXR emission is mainly from 
the LT region and the loop legs \citep{Wheatland95, Veronig04}.
There also exist a few large $\gamma$-ray flares, 
in which the LT bremsstrahlung source is detected up to 200--800 keV
during the decay of the HXR and $\gamma$-ray emission
and even has harder spectra than the FP sources \citep{Krucker08b},
interpreted as being due to long time trapping and collisional energy loss 
of the high energy accelerated electrons in the corona.

Simultaneous analysis of the LT and FP sources extending to high energies 
would be indispensable for a thorough understanding of 
acceleration and transport mechanisms.
For example, in the stochastic acceleration model, where
electrons undergo simultaneous acceleration and pitch angle scattering 
by plasma waves or turbulence in the coronal radiation region,
the spectral difference between the LT and FP sources 
can serve to determine the energy dependence of the escape time 
and pitch angle scattering time of the accelerated electrons
and thus better constrain theoretical models \citep{Petrosian99, Petrosian10}.
This requires high spatial and spectral resolution observations 
over a wide energy range of both the LT and FP sources. 
The {\it Yohkoh}/HXT has only four broad energy bands spanning from 14--93 keV, 
which greatly limits accurate determination of the spatially resolved spectra. 
For example, 
the 1991 January 13 flare \citep{Masuda94} has had a significant impact
on solar flare research \citep[see review in][]{Fletcher99, Krucker08a},
but to this date the nature of the spectrum 
of its coronal source still remains somewhat controversial 
\citep[e.g.][]{Masuda94, Alexander97, Masuda00, LiuR10}.
{\it RHESSI} provides HXR imaging spectroscopic observations with a higher 
resolution and sensitivity extending over a wider energy range 
to study the fundamental physics of energy release and particle acceleration
in solar flares \citep{Lin02}.
This allows for more accurate determination of the spectra of individual 
HXR sources and better understanding of the underlying physics. 
{\it RHESSI} observations have clearly shown nonthermal power law spectra
from the coronal sources and thus resolved the above controversy 
\citep[e.g.][]{Krucker08a, Krucker10}.

The aforementioned LT and FP sources are associated with a population of 
electrons propagating downward along a closed loop below the current sheet 
and constitute the majority of the flare HXR emission.
The bipolar X-type reconnection model also suggests 
existence of another electron beam above the current sheet,
which may propagates upward along open field lines through the corona
\citep[e.g.][]{Sturrock66, Aschwanden02}.
A second coronal HXR source,
which appears to be located above the LT coronal source, 
has been detected recently by {\it RHESSI} up to $\sim$20--30 keV 
in a few events \citep[e.g.][]{Sui03, Sui04, Veronig06, LiYP07, LiuW08}.
These two coronal sources exhibit an opposite spatial gradient,
for which the lower (upper) source appears at a higher (lower) altitude
with increasing energy, indicating a current sheet formed in between.
Such observations provide further evidence for 
magnetic reconnection and particle acceleration in solar flares.

In this paper, 
we present {\it RHESSI} imaging and spectroscopic observation 
of very energetic coronal HXR sources up to $\sim$100--150 keV 
simultaneously with two FP sources in a solar flare on 2003 November 3
(Solar Object Locator: SOL2003-11-03T09:43).
This is one of a few flares in which we found simultaneous LT and FP
sources above 50 keV observed by {\it RHESSI} during the impulsive phase
\citep{Chen09}.
In Section \ref{hecs} we present the imaging results of the coronal HXR sources.
In Section \ref{imsp} we mainly study the evolution of imaging spectroscopy of 
the LT coronal source and its comparison with the FP sources.
In Section \ref{disc} we discuss the implications arising from the above results
and a possible connection between the coronal HXR sources
and the type III radio bursts observed during the impulsive phase of the flare.
Finally, in Section 5 we briefly summarize our results.

\section{High Energy Coronal Sources}\label{hecs}

The 2003 November 3 solar flare under study is an intense eruptive event 
occurring in NOAA Active Region 10488 (N08\degr, W77\degr). 
According to the {\it GOES} soft X-ray profiles, 
the flare starts at 09:43 UT, peaks at 09:55 UT, and ends at 10:19 UT, 
and is classified as an X3.9 event.
This flare is accompanied with type III radio bursts
and a coronal mass ejection event \citep{Dauphin05},
but no solar energetic particles.
Earlier analysis of the flare identified a pair of coronal sources 
below $\sim$30 keV and two conjugate FPs based on the {\it RHESSI} HXR images 
reconstructed with the Clean method \citep{LiuW04, LiuW06, Veronig06}.
We show below that the two coronal sources actually extend to 
much higher energies by means of different image reconstruction methods.
These high energy coronal sources yield significantly 
new information about the coronal radiation regions. 

\begin{figure}[ht]
\epsscale{1.0}\plotone{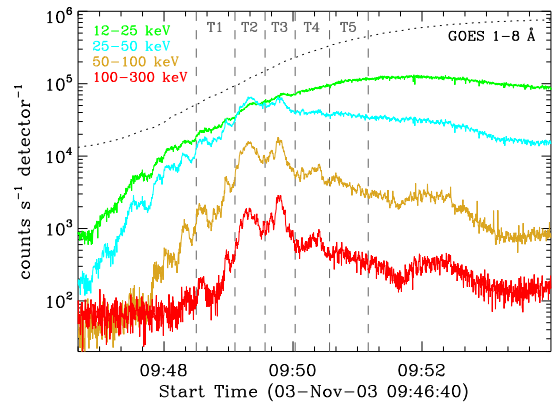}
\caption{Demodulated {\it RHESSI} count rates 
with 0.25 s resolution at four broad energy bins from 12 to 300 keV, 
superposed with the {\it GOES} soft X-ray flux at 1--8 \AA\
(dotted, in arbitrary units). 
The vertical lines (dash) delimit five time intervals for 
imaging and spectroscopic analysis, the third of which is the nonthermal peak.}
\label{fig_count}
\end{figure}

\begin{figure*}
\epsscale{1.15}\plotone{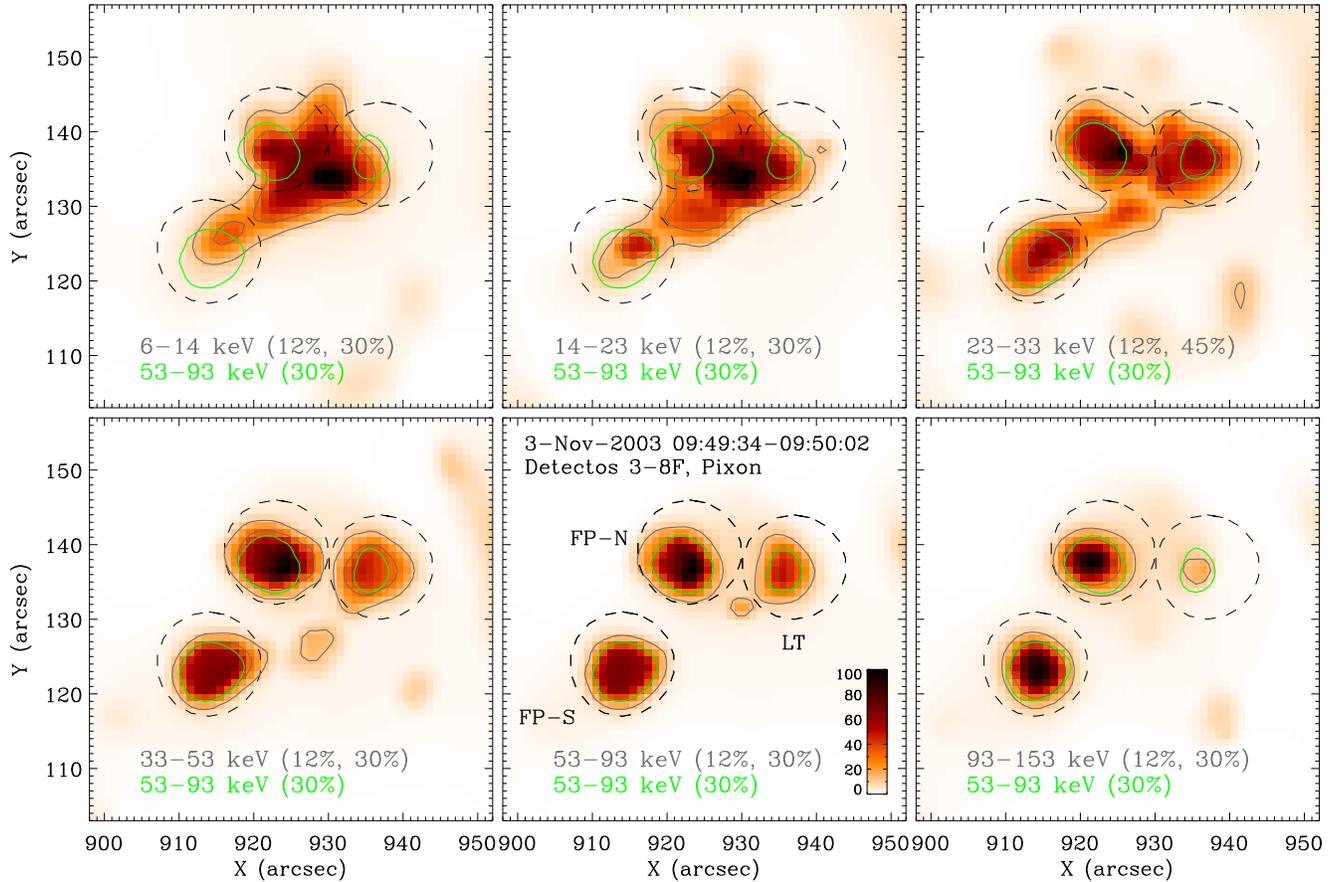}
\caption{HXR images at six broad energy bins from 6 to 153 keV
during the nonthermal peak generated by Pixon from the front segments 3--8
of FWHM $\sim$6.8\arcsec. 
Also shown are two contour levels of each image (gray) 
and the 30\% level of the 53--93 keV image (green).
The four intermediate energy bins from 14 to 93 keV
are designed to resemble the {\it Yohkoh}/HXT energy bands.
The images exhibit a loop structure below 23 keV,
and two FP sources and one LT coronal source dominating at higher energies.
The LT source at 33--153 keV 
falls within the $\sim$12\% level of the thermal loop emission.
Three circles (dash) with identical radii of 7\arcsec\ 
mark the regions from which HXR fluxes are extracted
for spectral analysis.
All images are scaled to the same color bar 
(bottom middle, in arbitrary units) for display.}
\label{fig_pixon6}
\end{figure*}

We focus on the impulsive phase of the flare (see Figure \ref{fig_count}).
The {\it RHESSI} count rates above 25 keV
show two main peaks around 09:49:20 and 09:49:48 UT,
while the rates at lower energies increase nearly monotonically.
In Figure \ref{fig_pixon6} 
we show the HXR images and source contours at six broad energy bins 
(6--14, 14--23, 23--33, 33--53, 53--93, and 93--153 keV) 
during the nonthermal peak 
as generated by the Pixon algorithm from the front segments 3--8.
The images from 6--23 keV show an asymmetric cusp-shaped loop structure
and faint, yet discernible emission from the southern FP of the loop.
At higher energies up to 93--153 keV, 
two FPs and one LT source dominate the HXR emission.
The $\ge$30 keV component of the LT source
is clearly revealed by the MEM\_NJIT algorithm and the Clean components as well,
but was missed in earlier analysis of the flare based on the Clean images.
We include comparison of these different algorithms in Appendix \ref{algo}.
We also discuss the pulse pileup effect in Appendix \ref{pileup} and conclude that
the LT source should not be due to the pileup effect.

\begin{figure}[t]
\epsscale{1.15}\plotone{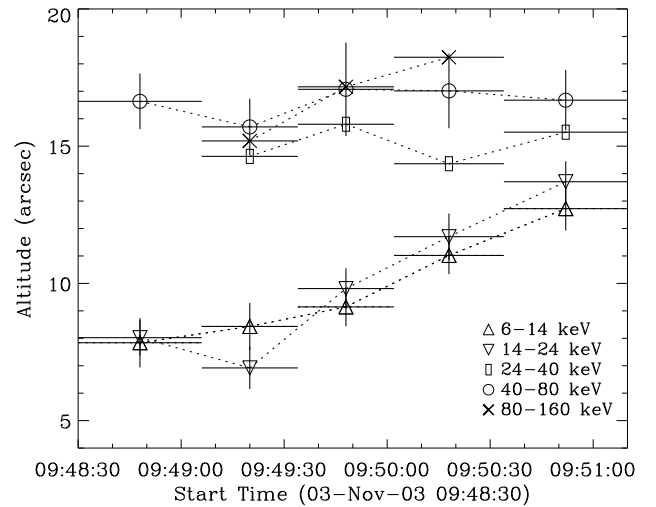}
\caption{Altitude of the LT centroid from 6--160 keV as a function of time.
Also shown are the error bars for the 6--14, 14--24, and 40--80 keV energy bins.
The uncertainty of the source centroid is estimated following
\citet{Bogachev05} and \citet{Mrozek06}, 
which for the current flare is approximated by the FWHM divided by 
the square root of the number of pixels within the 50\% contour level.}
\label{fig_altitude}
\end{figure}

\begin{figure*}[t]
\epsscale{1.15}\plotone{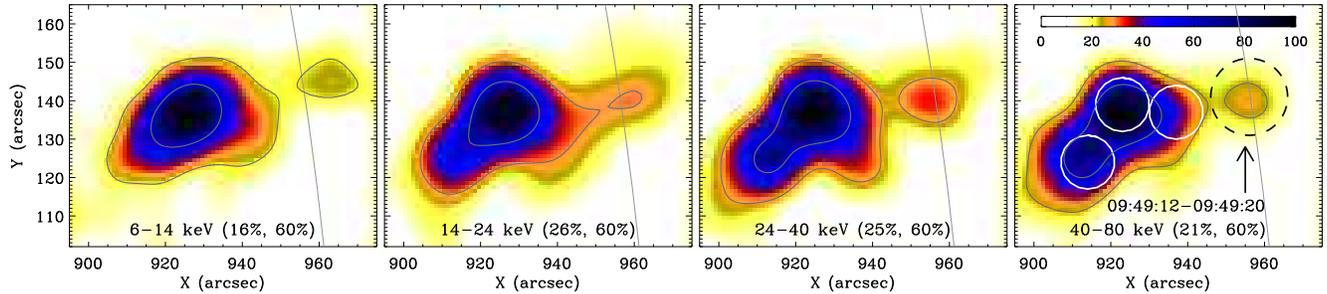}
\caption{Upper coronal source seen at four energy bins 
from 6 to 80 keV during part of the first peak (09:49:12--09:49:20 UT),
as indicated by the upward arrow and a circle of radius 10\arcsec\ (dash) 
in the rightmost panel.
The images are generated by the Clean algorithm (with natural weighting)
from the front segments 3--8,
superposed with two contour levels of each image.
The color bar is customized to highlight the upper coronal source.
The three smaller circles mark the LT and two FP sources 
seen in the Pixon images (see Figure \ref{fig_pixon6}).
The solid curve denotes the solar limb.}
\label{fig_cs}
\end{figure*}

\begin{figure*}
\epsscale{1.15}\plotone{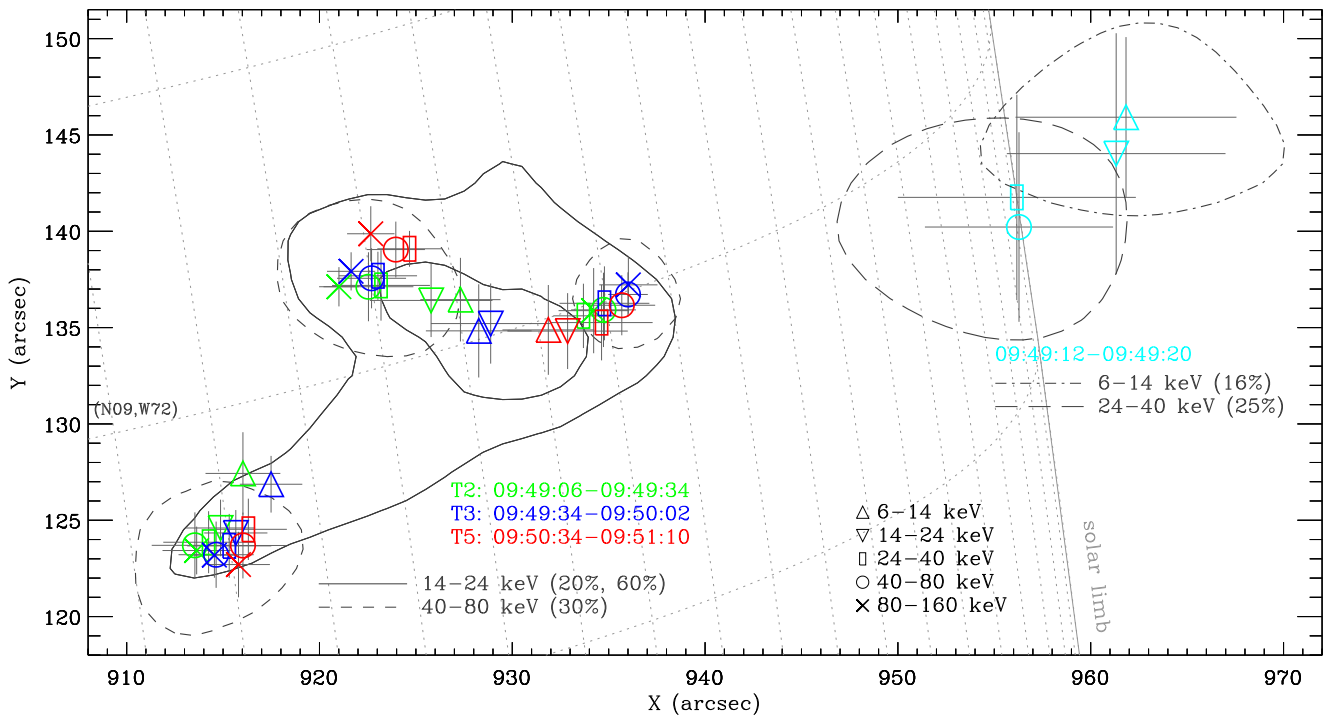}
\caption{Centroid distribution of individual HXR sources 
during the impulsive phase.
The left part shows the centroids of the LT and FP sources from the Pixon images
at five broad energy bins from 6--160 keV in three time intervals, 
superposed on the 14--24 keV (solid) and 40--80 keV (dash) contours 
from the nonthermal peak.
The LT centroids are mainly distributed along the circle of latitude at N09\degr.
The right part shows the centroids of the upper coronal source 
from the Clean images (with natural weighting) during part of the first peak,
supposed on the 6--14 keV (dash dot) and 24--40 keV (long dash) contours.
The centroids are calculated as the source position weighted with 
the intensity within the $\sim$50\% contour level of each source. 
The horizontal and vertical bars show the standard deviations of the centroids.
The dotted lines show the heliographic grids with 1$\degr$ separation.} 
\label{fig_cent}
\end{figure*}

This high energy LT source 
occurs nearly throughout the 2003 November 3 solar flare
and is most prominent during the impulsive phase. 
Its maximum intensity is about $\sim$50\% of the maximum FP intensity at 
33--93 keV and $\sim$20\% at 93--153 keV. 
As shown in Section \ref{imsp}, 
the LT flux at 50 keV is $\sim$4.5 photos s$^{-1}$ cm$^{-2}$ keV$^{-1}$,
comparable to the typical FP flux of an X1 class flare
\citep[see Figure 9 of][]{Saint-Hilaire08}.
Furthermore, the distinct LT source above 23 keV 
clearly lies above the most intense part of the thermal loop
and is well separated from the FP sources.
The projected angular separation between the centroids of 
the high energy LT and the thermal loop is $\sim$8\arcsec.
This unusual LT source observed by {\it RHESSI} is reminiscent of 
the ``above-the-loop-top" coronal source seen up to 33--53 keV 
in the 1991 January 13 flare observed by {\it Yohkoh}/HXT,
which lies above 
the thermal soft X-ray loop by $\sim$10\arcsec\ \citep{Masuda94}.
In comparison,
the nonthermal and thermal components of the coronal sources 
are mostly cospatial as observed in partially occulted flares 
\citep{Tomczak01, Tomczak09, Krucker08c}.
On the other hand, despite its large separation from the 
intense thermal loop, this high energy coronal source
should still be located at the top of some 
cusp-shaped magnetic loop, which may not yet become fully visible 
at soft X-rays \citep[see also][]{LiuW08, Longcope11}.
Therefore, in this paper, 
we use the term ``loop top (LT)" for this coronal source.

Earlier analysis of the flare showed
that the thermal LT moves upward (after an early altitude decrease) and 
the two FPs move apart as the flare proceeds in time \citep{LiuW04, Veronig06},
supporting the standard magnetic reconnection model \citep[e.g.][]{Priest00}.
Now we investigate the spatial distribution of the LT source at different energies.
In Figure \ref{fig_altitude} we plot the altitude of the centroids of the LT source
within its $\sim$50\% contour level
at five broad energy bins (6--14, 14--24, 24--40, 40--80, and 80--160 keV) 
in five time intervals during the impulsive phase (see Figure \ref{fig_count}). 
First, in general the centroids appear at higher altitudes
with increasing energy (see also Figure \ref{fig_cent})
and the centroids above 24 keV show a displacement of $\le$8\arcsec\ 
from those at lower energies.
Second, the LT centroids below 24 keV move gradually to 
higher altitudes (up to $\sim$6\arcsec) 
with a velocity $\sim$30 km s$^{-1}$ as the flare develops.
While at higher energies, the LT source shows little motion except at 80--160 keV.
These behaviors indicate
very efficient confinement of the accelerated electrons in the solar corona.

Furthermore, 
we show in Figure \ref{fig_cs} that the second (upper) coronal source sitting 
around the west solar limb \citep{Veronig06} actually has significant emission 
even at energies up to $\sim$80 keV.
This high coronal source can be best detected in the Clean images 
around the first peak of the flare and with a relatively short integration time.
It appears rather distinctive and separated from 
the underlying closed loop consisting of the LT and FP sources.
Its altitude is roughly 2--3 times that of the high energy LT source.
We plot in Figure \ref{fig_cent} the spatial distribution of its centroids,
along with the LT and FP centroids.
The energy dependence of its centroids exhibits nearly an opposite trend 
compared to the LT; the higher energy this coronal source, 
the lower its altitude toward the flare loop.
In other words, the centroids of both coronal sources are closer to 
the imagined reconnecting X-point at higher energies.
This flare provides an example of a current sheet 
(of an extent $\sim$20\arcsec) as inferred from X-ray
observation of the outflow regions 
with much higher energies than other events. 

\section{Imaging Spectroscopy}\label{imsp}

\begin{figure*}
\epsscale{1.15}\plotone{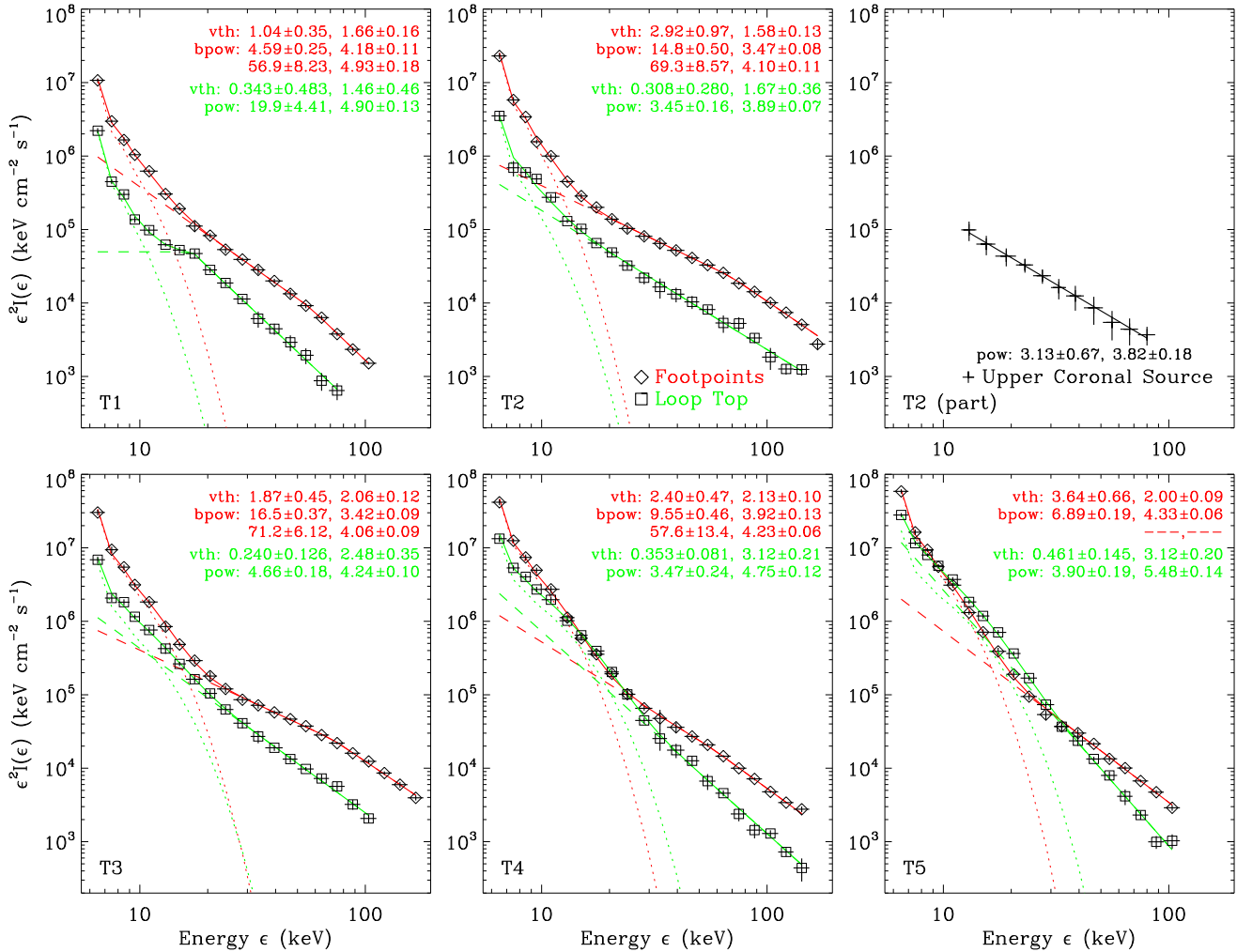}
\caption{HXR spectra $I(\epsilon)$ from the LT source ({square, green})
and the two FP sources summed ({diamond, red}) 
in five time intervals (see T1--T5 in Figure \ref{fig_count}).
The spectra are fitted by 
an isothermal bremsstrahlung spectrum plus a single or broken power law
from 6 keV to the highest energies at which the sources are visible.
For the first interval, we also include a low energy break at $\sim$20 keV.
As for the fitting parameters, 
the isothermal function (``vth") includes the emission measure (${\rm EM}$, 
in units of 10$^{49}$ cm$^{-3}$) and temperature ($T$, keV);
the single power law (``pow") includes the normalization of the spectra
at 50 keV and the spectral index $\gamma$;
the broken power law (``bpow") further includes 
the break energy and the index above the break.
Note that below $\sim$20 keV, the LT spectra 
have significant contamination from the thermal loop emission and thus 
the displayed thermal parameters do not faithfully represent the LT source;
also the northern FP source partly overlaps the loop emission, thus 
the summed FP flux is overestimated.
The top right panel shows the spectrum 
from the upper coronal source (plus) during 09:49:12--09:48:24 UT
and the corresponding power law fitting.
In order to highlight the difference between the steep LT and FP spectra,
we have multiplied $I(\epsilon)$ with $\epsilon^2$ for display.
This representation also indicates the photon energy at which
most of the energy is radiated.}
\label{fig_ospexspec}
\end{figure*}

In this section we present results from 
imaging spectroscopic analysis of individual HXR sources,
which is implemented using the Object Spectral Executive 
\citep[OSPEX; ][]{Smith02} package of the Solar SoftWare (SSW). 
We use the Pixon algorithm to reconstruct images from 6 to $\sim$180 keV. 
We then extract the spatially resolved HXR spectra 
(photons cm$^{-2}$ s$^{-1}$ keV$^{-1}$) 
from the LT and FP sources
over three fixed circles with radii 7\arcsec\ (see Figure \ref{fig_pixon6}).
Following the procedure currently adopted 
in OSPEX for error estimate \citep[see e.g.][]{Saint-Hilaire08},
we take one third of the maximum flux outside the flaring region 
to be the 1-$\sigma$ uncertainty of the flux of each source.
Finally we fit the HXR spectra parametrically to the combination of 
an isothermal bremsstrahlung spectrum 
and a single or broken power law \citep{Holman03}
using {\it RHESSI}'s full spectral response matrix \citep[e.g.][]{LiuW08}.

\subsection{HXR Spectra}
In Figure \ref{fig_ospexspec}, we show the HXR spectra $I(\epsilon)$ 
for the LT and the summed FPs and the corresponding fitting.
The LT spectra can be well fitted by a thermal function plus a power law tail
with an index $\gamma\equiv -d\ln I(\epsilon)/d\ln\epsilon$ 
varying from $\sim$4--5.5. 
If we take the emission measure ${\rm EM} = 0.3 \times 10^{49}$ cm$^{-3}$
and the size $L=10^9$ cm, 
and assume a filling factor of unity, we obtain 
a density ${n}=\sqrt{{\rm EM}/L^3}\simeq 5\times10^{10}$ cm$^{-3}$
averaged over the circle ``LT" in Figure \ref{fig_pixon6}.
Here we should note that since
the circle ``LT" includes part of the low energy thermal loop and
the neighboring pixels reconstructed from indirect Fourier imaging
(as employed by {\it RHESSI}) are not independent, 
the thermal emission within this circle mainly comes from 
part of the thermal loop.
Therefore the thermal properties of the high energy LT source
cannot be well measured and 
the above value $n$ may only give an upper limit for the LT density,
which could be significantly lower.
In comparison, the average density of the entire thermal loop is 
estimated to be $\sim$$8\times10^{10}$ cm$^{-3}$. 

In Figure \ref{fig_ospexspec} top right panel, 
we also plot the spectrum of the upper coronal source
that is located above the LT source during the first peak of the flare.
It turns out that the two coronal sources have comparable spectral indices.
The similarity in HXR spectra may provide further evidence that
the same acceleration mechanism is responsible for the two oppositely
directed electron beams that generate the two coronal HXR sources. 

In contrast, the FP spectra above $\sim$20 keV 
are stronger and flatter and are better fitted by a broken power law.
At lower energies, 
the FP spectra show a softer, perhaps quasi-thermal component
which can be fitted by a thermal function of a temperature $\sim$$2\times10^7$ K.
In particular, the southern FP source, 
which has little contamination from the intense loop emission 
(see Figure \ref{fig_pixon6}), 
clearly shows such a prominent thermal-like component.
Thermal X-ray emission from the FP regions has been detected 
from a few other flares \citep[e.g.][]{McTiernan93, Hudson94, BB06}
with a temperature as high as $\sim$$10^7$ K \citep{Hudson94}.
The origin of the thermal-like component of the FP spectra 
in the current flare is puzzling (see also Section \ref{ss}).

\begin{figure}
\epsscale{1.15}\plotone{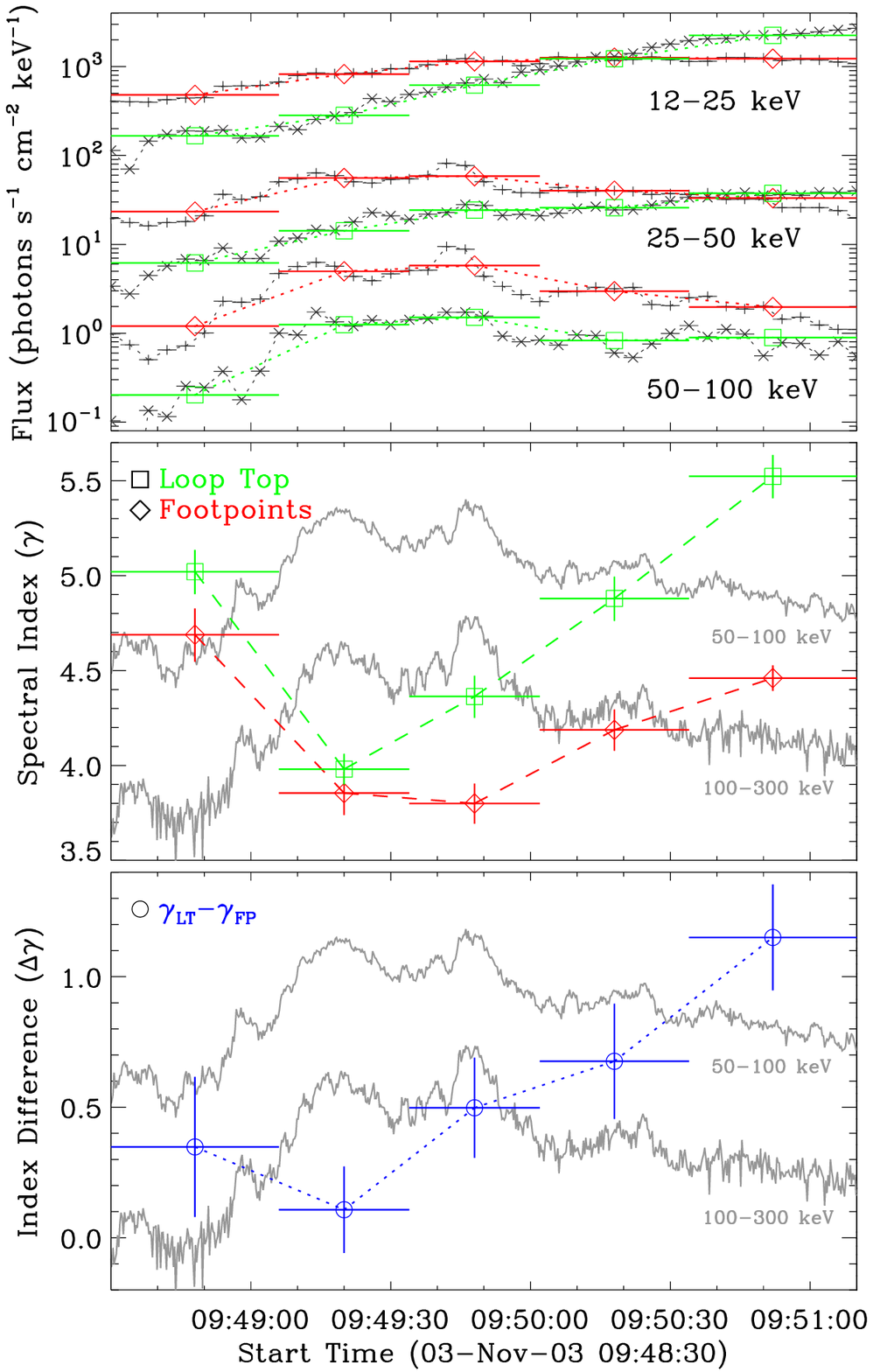}
\caption{Spectral evolution during the impulsive phase.
Top: HXR fluxes from the LT source ({square, green}) 
and the two FP sources summed ({diamond, red})
at 12--25, 25--50, and 50--100 keV.
Also shown are the 4 s resolution data (black).
The fluxes are extracted from the Pixon images.
Middle: (mean) power law indices of the LT and the summed FP spectra.
{Bottom}: difference in power law indices (circle, blue) 
between the LT and the FPs.
The lower two panels are superposed on the spatially integrated fluxes 
at 50--100 and 100--300 keV (in arbitrary units).}
\label{fig_ospextime}
\end{figure}

\subsection{Spectral Relation and Evolution}

Now we investigate the temporal evolution of the LT and FP spectra
and their relative difference as shown in Figure \ref{fig_ospextime}.
First, in the top panel, 
we show the time profiles of the fluxes from the LT and the summed FP sources
with a time resolution of 4 seconds.
Below 25 keV, both the LT and FP fluxes show gradual increase from the beginning;
while above 25 keV, they show impulsive profiles that are well correlated
but there is also some difference.
Also note that the ratio between the LT and the FP fluxes increases with time
(i.e. the LT becomes relatively more prominent compared to the FPs).
This could be partly due to 
the successive formation of larger coronal loops as reconnection proceeds 
and the effect of chromospheric evaporation,
which fills the coronal loop with dense chromospheric plasma,
making bremsstrahlung emission brighter around the LT region.

Second, as in the middle panel,
the spatially resolved LT and FP spectra 
roughly exhibit soft-hard-soft evolution, namely, 
the spectral index is anti-correlated with the flux during the flare.
Note that for the FP spectra, 
we take the average value of the two indices below and above the break energy.
We have also fitted the LT and FP spectra above 50 keV by a power law,
as similarly done by \citet{Emslie03} and \citet{LiuW09a}, 
and found a similar trend as above.
This behavior has been found in 
the spatially integrated spectra \citep[e.g.][]{Grigis04}
or spatially resolved spectra from {\it RHESSI} \citep[e.g.][]{BB06}. 
Such spectral evolution may reflect an intrinsic property of 
the acceleration mechanism and can be explained by 
the stochastic acceleration model
\citep[e.g.][]{Petrosian04, Grigis06, Bykov09, LiuS09}.
However, there is no exact peak to peak correspondence seen in this flare.

Third, as in the bottom panel, 
the difference between the LT and the FP spectral indices, 
$\Delta\gamma =\gamma_{\rm LT}-\gamma_{\rm FP}$, 
is very small during the impulsive phase. 
It is smaller than or around one, 
being around zero during the first peak.
We have also fitted the ratio between the LT and FP spectra 
by a power law and found a similar index difference.
The difference here is comparable to or even smaller than 
the most extreme cases found in previous statistical study 
for the {\it Yohkoh} flares \citep{Metcalf99, Petrosian02}
and {\it RHESSI} flares \citep{BB06, LiuW06, Shao09}.
It is also considerably smaller than the average spectral difference between 
the (LT) coronal emission from partially occulted flares 
and the FP emission from non-occulted flares,
$\sim$1.5 for the {\it Yohkoh} flares \citep{Tomczak09} and
$\sim$2 for {\it RHESSI} flares \citep{Krucker08a, Krucker08c}.
The observed small spectral difference reflects why the 2003 November 3 flare
has very bright LT emission up to $\sim$100--150 keV 
that can still be detected simultaneously 
with the intense FPs under the current dynamic range $\sim$10.

\section{Discussions}\label{disc}

The imaging and spectroscopic analysis of the coronal HXR sources 
in the 2003 November 3 flare observed by {\it RHESSI} strongly indicate that 
electron acceleration is closely related to the energy release process
through the reconnecting current sheet in solar flares. 
The spatial variation of the LT coronal source 
may be indicative of the cusp-shaped geometry of 
the outflow region below the current sheet.
Detection of coronal sources extending to high energies 
can severely challenge theoretical models and 
constrain intrinsically complex physical processes in solar flares. 
In this section, we first discuss a few aspects mainly arising from 
the newly detected high energy LT coronal source and compare them with
some previous models, in particular, the stochastic acceleration model.
Second, we attempt to explore a possible connection between 
the coronal HXR sources and the type III radio bursts observed 
during the impulsive phase of the flare.

\subsection{Spatial Structure}\label{ss}

Existence of the distinct coronal sources up to $\sim$100--150 keV 
clearly implies that the electrons are accelerated to $\ge$200--300 keV and 
are confined near these sources.
This may result from pitch angle scattering of electrons
by turbulence, as advocated by 
the stochastic acceleration model \citep[e.g.][]{Petrosian99, Petrosian04}.
The opposite spatial gradient of two coronal sources 
observed in this and a few other events, in which 
the sources get harder toward the presumed reconnection X-point,
can also be explained by this model \citep{LiuW08}.
In addition, the electron escape time from the acceleration site in this flare 
increases with energy \citep{Petrosian10}, effectively enhancing their confinement
(see, however, the discussion below). 
In this and other models the electrons escaping from the LT coronal region
will produce HXR emission at the thick target FP regions
and drive evaporation flow to fill the coronal loop, 
while those electrons escaping from the upper coronal source may be responsible 
for the type III radio emission (see Section \ref{iii} below).

Another model often advocated for trapping of electrons in the corona 
is magnetic field convergence below the current sheet \citep{Fletcher98}.
Such a magnetic bottle can generate a distinct LT coronal source,
but not the variation of the source centroid with energy.
Recently, a drift-kinetic model 
including betatron acceleration and collisional pitch angle scattering 
during field line shrinkage \citep{Minoshima11} shows that 
the height of the accelerated electrons in the corona increases with energy 
up to a few tens of keV, but then decreases at higher energies.
This may qualitatively explain previous observations
of the height distribution of low energy coronal sources,
but it cannot account for the increase of the coronal LT height 
at higher energies up to $\sim$100--150 keV 
(with corresponding electron energy $\ge$200--300 keV)
as observed in the current flare.

Ever since the observation of the above-the-loop-top source 
in the 1991 January 13 flare \citep{Masuda94}, it is often stated that 
electron acceleration takes place above the thermal soft X-ray loop. 
However, the spatial variations described above indicate that 
the situation is more complex. 
Because the more abundant lower energy electrons are located at 
lower (or inner) field lines and because they are the more effective agent 
of heating and evaporation compared to the high energy electrons,
it may be natural to expect the thermal loop 
to lie somewhat below the LT coronal HXR source.
More thorough modeling including transport of electrons 
in an inhomogeneous environment is required to address these details.

Finally, we briefly discuss the spatial distribution of the FP sources. 
They in general appear deeper at the lower part of the loop
with increasing energy (see Figures \ref{fig_pixon6} \& \ref{fig_cent}),
which at first sight seems consistent with the classical thick target model
for electron transport.
However, the existence of the low energy southern FP source at $\sim$6 keV
is difficult to explain when considering that 
the column density along the loop leg 
$\sim$$8\times 10^{19}$ cm$^{-2}$ (see Section \ref{imsp}) 
can collisionally stop electrons with energy%
\footnote{In the classical thick target model \citep[e.g.][]{Brown02}, 
a column density ${\cal N}$ can stop (nonrelativistic) electrons with energies 
up to $E=7.2~({\cal N}/10^{19}~{\rm cm}^{-2})^{1/2}$ \ {\rm keV},
if the Coulomb logarithm $\ln\Lambda=20$.}
up to 20 keV. 
Although quasi-thermal electrons may escape from the corona
as shown in the stochastic acceleration model \citep{Petrosian04},
the above collisional effect basically leads to
little low energy thermal emission at the FPs \citep[e.g. Figure 12 of][]{LiuW09b}.
In addition, the gradual rising time profiles of the FP emission at such low energies
do not seem to agree with the scenario of direct heating  
due to nonthermal electrons at the FP regions \citep[][]{Hudson94}.
Therefore the origin of the low energy FP sources of quasi-thermal spectra 
observed in this flare is not clear.
It deserves further study in the future whether the current observation 
requires more valid transport models for energetic electrons escaping from the corona 
or implies some new energization processes in the dense chromosphere
\citep[e.g.][]{Fletcher08, Brown09}.

\subsection{Spectral Properties}

The 2003 November 3 flare shows the usual soft-hard-soft evolution of 
the LT and FP spectra (Figure \ref{fig_ospextime}, middle panel). 
This is a natural consequence of the stochastic acceleration model,
where the spectral hardness of the accelerated electrons, 
and consequently that of the emitted HXR photons
increase primarily with the level of turbulence intensity 
\citep[e.g.][]{Petrosian04}. 
A higher level of turbulence may also cause more efficient trapping, 
which is equivalent to slower escaping of the accelerated electrons
to the FPs of the loop. 
This may be the explanation for the evolution of the LT and FP 
spectral difference (Figure \ref{fig_ospextime}, bottom panel).

However, the most unusual spectral aspect of this flare is 
the observation of the LT source up to high energies ($\sim$100--150 keV),
which is made possible because of its relatively hard spectrum. 
As emphasized above, the observed spectral index difference between 
the LT and FP sources ($\Delta\gamma \simeq$~0--1) is significantly 
smaller than those found in the more frequent, less intense flares. 
Quantitatively, 
this spectral difference is determined by 
the energy dependence of the escape time in the acceleration region,
which is related to the pitch angle scattering rate.
As derived directly from the electron flux images, 
the escape time from the LT acceleration site during the nonthermal peak
of the flare increases with energy \citep{Petrosian10}, 
requiring a scattering time that is shorter than 
the crossing time and decreases with energy relatively rapidly. 
This does not seem to have a simple explanation 
considering the models proposed so far for the LT coronal emission.

First, we note that in previous literature for stochastic acceleration
\citep[e.g.][]{Miller90, Pryadko97, Pryadko98, Chandran03, Emslie04, 
Petrosian04, Grigis06, Bykov09, LiuS09},
the electron escape time is either simply taken to be the same as 
the crossing time $\tau_{\rm cross}=L/v\propto 1/\sqrt{E}$, 
which is the case when the scattering time 
$\tau_{\rm scat} \gg \tau_{\rm cross}$, 
or set to $\tau_{\rm cross}^2/\tau_{\rm scat}$ for the opposite case. 
The scattering time can be calculated numerically assuming 
a Kolmogorov-type spectrum of turbulence and/or 
parallel propagating plasma waves along magnetic field lines.
The latter also gives rise to an escape time that is 
very flat or decreases with electron energy \citep[e.g.][]{Petrosian04}. 
The escape time of such an energy dependence roughly leads to 
$\Delta\gamma\simeq$ 1.5--2 between the LT and FP spectra
in the energy range of interest for HXR observations. 

One of the earliest models suggested for production of a distinct LT source 
is collisional confinement by a dense region in the corona \citep{Wheatland95}.
In this model, the high energy LT source is thin target in nature and its spectrum is 
softer than the thick target FP spectrum by $\Delta\gamma\simeq 2$. 
One may include transport effects from the corona to the chromosphere,
such as return current energy loss \citep[e.g.][]{Zharkova06, BB08},
a nonuniform ionization target \citep[e.g.][]{Su09},
and wave-particle interactions \citep[e.g.][]{Holman82, Hannah11},
to overcome the difficulty. 
However, all these effects in general tend to 
make the electron spectrum flatter and thus cause $\Delta\gamma > 2$.

One model based on collisional pitch angle scattering of electrons 
is that of converging magnetic field lines in the corona \citep[e.g.][]{Fletcher98}. 
This may effectively increase the electron escape time with energy
to be $\propto E^{3/2}$ \citep[see][]{Melrose76}. 
However, as shown in \citet{Petrosian10},
the required escape time varies less rapidly with energy than $E^{3/2}$,
and more importantly, it is about ten times smaller than 
the collisional energy loss time for this flare. 
In addition, this effect alone tends to 
make the coronal LT spectrum progressively harder with time 
and the LT flux decay slower than the FP flux,
which contradict what are observed during the impulsive phase
(see Figure \ref{fig_ospextime}).

Therefore, it seems that in this special flare one needs to more effectively 
confine the high energy electrons in the acceleration region than predicted 
by the above models.
We believe that the observed small spectral difference 
during the impulsive phase should most likely be related to 
some unique conditions or the acceleration or transport processes in the corona.
For example, as pointed out in \citet{Petrosian10}, 
one possibility to increase the escape time for high energy electrons
is to invoke a turbulence spectrum 
that is steeper than the commonly assumed Kolmogorov-type spectrum,
as is expected for damped turbulence beyond the inertial range.
Another possibility is 
acceleration by perpendicularly propagating plasma waves, 
which can preferentially accelerate electrons with the pitch angle near 90\degr\ 
\citep[e.g.][]{Petrosian99}. 
Compared to the unidirectional beam distribution or isotropic distribution,
such a pancake-like distribution can more easily confine electrons in the corona.
Finally, it is likely that rather than operating separately,
some of these conditions may be operating simultaneously 
and be capable to account for the observed spectral difference and 
its evolution. 
A combination of turbulence scattering and magnetic field convergence 
may possibly yield an escape time increasing with energy, 
yet much shorter than the collisional loss time.

\begin{figure}
\epsscale{1.15}\plotone{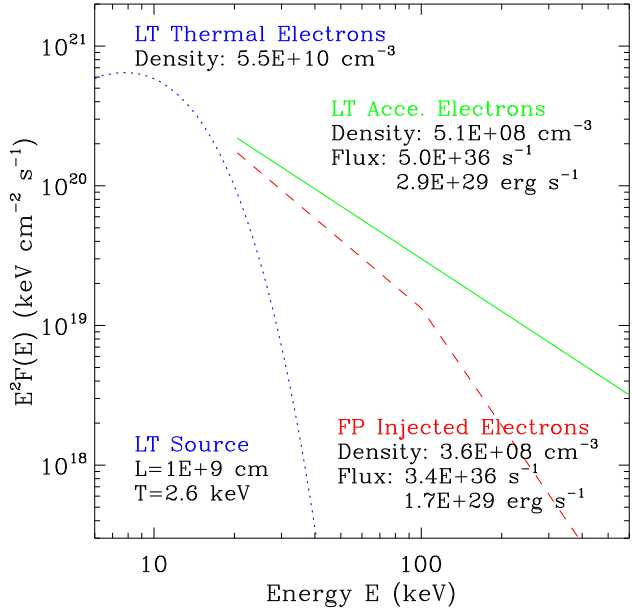}
\caption{Flux spectra $F(E)$ of the accelerated electrons
as inferred from the thin target LT bremsstrahlung spectrum 
(for a background density $\sim$$5\times10^{10}$ cm$^{-3}$) 
and the escaping (injected) electrons from the thick target FP spectrum
during the nonthermal peak (28 seconds) by parametric forward fitting.
The LT electron spectrum can be well fitted by a power law (solid, green)
with $\delta\simeq3.3$ and $E_{\rm c} = 20$ keV.
The FP electron spectrum can be better fitted 
by a broken power law (dash, red) with the break energy set at 100 keV
and an index $\sim$3.6 and $\sim$4.8 below and above the break, respectively.
Also shown is the thermal electron distribution at the LT source (dotted, blue).
Note that the prescribed two-component model in the forward fitting procedure
does not guarantee a continuous transition of the electron spectra.}
\label{fig_elec}
\end{figure}

\subsection{Thermal and Nonthermal Electrons}

Conventionally, the thick target bremsstrahlung of 
nonthermal electrons injected into the loop is used to model 
the spatially-integrated HXR emission from the flare loop.
While for the current flare under study,
imaging spectroscopic observation of the spatially resolved LT and FP sources
over a range $>$100 keV provides a unique opportunity to infer the processes
of electron heating, acceleration, and escaping in the corona.
Assuming that the LT source is resulting from the thin target bremsstrahlung
and the FP sources are due to thick target emission by the escaping electrons,
both with an addition of a thermal component,
we adopt the forward fitting method \citep{Holman03} 
to independently infer the accelerated electron flux spectrum at the LT
and the escaping (or injected) electron flux spectrum to the FP sources.

As shown in Figure \ref{fig_elec}, during the flare nonthermal peak, 
the electron flux spectrum at the LT can be well fitted by a power law
with an index $\delta\simeq3.3$ and a low energy cutoff $E_{\rm c} = 20$ keV, and
the (instantaneous) nonthermal electron density is $\sim$$5\times10^8$ cm$^{-3}$
assuming a background density $\sim$$5\times10^{10}$ cm$^{-3}$ (see Section \ref{imsp}).
While the injected spectrum to the FPs can be better fitted by a broken power law,
which is steeper and weaker than the LT accelerated spectrum.
This difference may result from 
that the low energy electrons escape more easily from the corona 
to the FP regions than higher energy electrons and 
that the escape time is longer than the crossing time 
\citep{Petrosian10}.

Given an upper limit of the background density $\sim$$5\times10^{10}$ cm$^{-3}$ 
and temperature $\sim$2.5 keV, 
the density and total energy of the accelerated electrons $\ge$20 keV
are roughly 1\% and 10\% (lower limits) of those of the background plasma, 
respectively.
Such numbers are representative of the impulsive phase.
It is worth mentioning that
the above estimate of the number percentage of the accelerated electrons
is inversely proportional to the background density squared or 
the emission measure at the nonthermal LT source.

On the contrary, 
if the LT density is indeed significantly lower than the above upper limit 
(see discussion in Section \ref{imsp}), then essentially 
all the background electrons would be accelerated into a power law.
This will be similar to recent measurements of 
the above-the-loop-top coronal sources
observed by {\it RHESSI} in a C8.3 class flare \citep{Krucker10} 
and an M9.9 class flare \citep{Ishikawa11b}.
Accordingly, 
the electron flux spectrum at the LT source (see Figure \ref{fig_elec}) 
will increase by $\sim$10 times compared to the original assessment,
so will the density of the accelerated electrons, 
which becomes to be $\sim$$5\times10^{9}$ cm$^{-3}$.

\subsection{Type III Radio Bursts and Coronal Sources}\label{iii}

\begin{figure}
\epsscale{1.15}\plotone{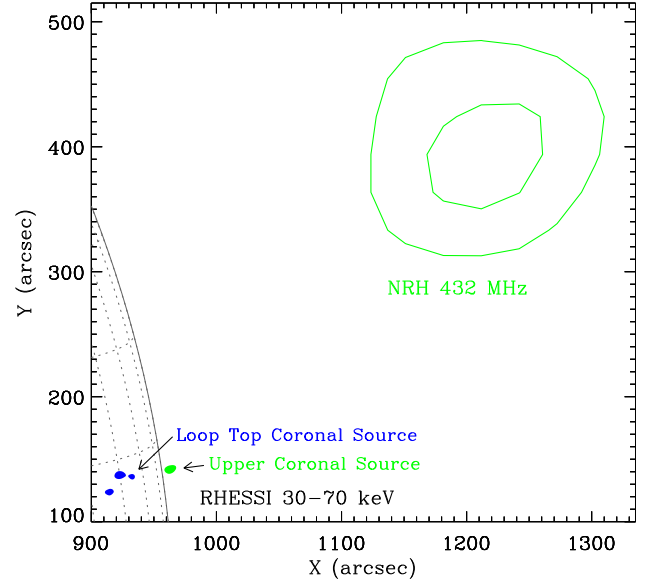}
\caption{NRH 432 MHz radio source (40\% and 80\% contour levels, green)
at 09:49:27 UT with an integration time of 0.9 s
and the {\it RHESSI} HXR sources at 30--70 keV from 09:49:24--09:49:32 UT.
Two arrows are drawn to point to the LT coronal source (blue) 
and the upper coronal source (green) indicating 
two electron populations that are spatially separated.
The dotted lines show the heliographic grids with 5$\degr$ separation.} 
\label{fig_radio}
\end{figure}

The flare accelerated electrons 
with access to open field lines eventually escape from the corona
and produce type III radio bursts, which are characterized as features 
rapidly drifting from high to low frequencies in radio spectrograms 
\citep[e.g.][]{Bastian98}.
The close temporal correlation between the type III bursts and the
spatially integrated HXR emission observed in many flare events 
suggests that the corresponding electron beams may come from 
a common injection site or result from the same acceleration process 
in the corona \citep[e.g.][]{Kane81, Aschwanden95}.
Recently a few solar eruption events with simultaneous observations of 
type III radio bursts and {\it RHESSI} HXR emission have been analyzed 
\citep[e.g.][]{Vilmer02, Krucker08d, Christe08, Bain09, Reid11}.
However, so far the relation between the electron populations producing
type III bursts and individual HXR sources still remains elusive.

Ground-based radio observation of the 2003 November 3 flare 
indicates that the impulsive phase 
is associated with many episodes of type III radio emission 
rapidly drifting from $\sim$400--500 MHz to lower frequencies
in the decimetric/metric range \citep[see][]{Dauphin05}.
The radio sources that are related to the type III bursts exhibit 
some spatial dispersion with frequency \citep{Dauphin05, Dauphin06}.
In Figure \ref{fig_radio}, we plot only 
the radio source at 09:49:27 UT from the highest frequency of 432 MHz
observed by the Nan\c{c}ay Radioheliograph \citep[NRH;][]{Kerdraon97}.
The local plasma density corresponding to the onset frequency 
of the type III bursts is estimated to be $\sim$$6\times10^8$ cm$^{-3}$,
if the plasma emission is at the second harmonic.
This density is about two orders of magnitude lower than 
that at the LT region (see Section \ref{imsp}).
By adopting a coronal density model for conditions above active regions 
which uses 10 times of the Baumbach-Allen formula \citep[e.g.][]{Paesold01},
such a density corresponds to a height of 
$\sim$0.23 $R_\sun$ above the solar surface,
very close to the observed location of the 432 MHz radio source.

Generation of type III radio bursts during the flare impulsive phase 
requires that 
the accelerated electrons have access to open field lines in the corona.
In the bipolar model, open field lines can only exist above 
the reconnecting X-point \citep[e.g.][]{Sturrock66, Aschwanden97, Aschwanden02}.
The flare accelerated electrons above the X-point 
that produce the upper coronal source eventually escape from the lower corona 
and induce the observed type III radio bursts (Figure \ref{fig_radio}). 
This flare provides a unique opportunity to connect 
the upper coronal HXR source and the type III radio bursts
through the electron population located above the X-point.

\section{Summary}

Finally, 
we briefly summarize our results of 
the newly found high energy coronal HXR sources 
in the 2003 November 3 solar flare as observed by {\it RHESSI}. 

1. The LT coronal source can be detected up to $\sim$100--150 keV and 
the upper coronal source up to $\sim$40--80 keV
located about 20\arcsec\ above the LT source, 
much higher than the energies commonly observed during the impulsive phase
when the intense FP emission is also present in the field of view.
The opposite spatial gradient of the two coronal sources
indicates that electron acceleration is intimately related to 
the reconnecting current sheet. 
The spectra from the coronal sources 
can be described by a power law of a similar index.
The high energy LT source exhibits an impulsive temporal profile 
and soft-hard-soft spectral evolution.
The electron density and the percentage of accelerated electrons
at the LT source are not well constrained, 
but a range between $\sim$(0.5--5)$\times10^{10}$ cm$^{-3}$ 
and $\sim$(1--100)\% can be obtained, respectively.
These spatial and spectral properties of the coronal sources 
qualitatively support the stochastic acceleration model.
In this scenario, 
two spatially separated populations of electrons are 
scattered and accelerated by plasma waves or turbulence 
below and above the current sheet,
generate the two distinct coronal sources by bremsstrahlung, 
and finally escape from the acceleration regions to the FP regions
of the flare loop or moving upward in the corona.
This latter electron population may further produce 
the type III radio bursts observed in the flare. 

2. The LT and FP spectral difference ($\Delta\gamma\simeq$ 0--1)
is found to be much smaller than commonly seen during the impulsive phase.
Such a difference should most likely be ascribed to 
the physical processes in the coronal radiation region.
The small difference requires a steeper spectrum of the escaping electrons
than that of the accelerated electrons, 
or an escape time increasing with electron energy.
In contrast, stochastic acceleration of electrons 
by turbulence of the usually assumed Kolmogorov-type spectrum produces
an escape time that is very flat or decreases with electron energy.
Therefore, more efficient pitch angle scattering and acceleration 
are required to explain the above observation.
{\it More realistic modeling including 
both kinetic effects and the macroscopic flare structure
is needed to address this issue. 
Inclusion of pitch angle anisotropy of electrons
and convergence of magnetic field lines in the corona
may help mitigate the above spectral discrepancy.}

The above results from imaging spectroscopic analysis of 
the spatially resolved sources in the 2003 November 3 solar flare 
highlight the importance of HXR observations
with a high dynamic range and sensitivity 
\citep[e.g. through focusing optics, see][]{Krucker11}
and over a wide energy range
in understanding electron acceleration (and transport) processes.

\acknowledgments
We thank the {\it RHESSI} team for providing the HXR data 
and the Nan\c{c}ay Radioheliograph team for providing the radio data.
We thank the referee S\"am Krucker for constructive comments
that helped improve the paper.
We thank Siming Liu for valuable discussions of the early results,
Richard Schwartz, Gordon Hurford, S\"am Krucker, David Smith,
and Wei Liu for discussions of the pulse pileup effect,
Nicole Vilmer and Steven White for discussions of the radio observation,
Kim Tolbert, Wei Liu, and Brian Dennis for discussions of 
OSPEX spectral fitting,
Gordon Hurford for discussions of image reconstruction algorithms,
Gordon Emslie and Anna Massone for demonstrating the uv\_smooth method,
and Pascal Saint-Hilaire for discussions of centroid uncertainty.
{\it RHESSI} is a NASA small explorer mission. 
This work was supported by NSF SHINE grant ATM0648750 and NASA grant NNX10AC06G.

{\it Facilities:} \facility{{\it RHESSI}}.

\appendix
\section{Comparison of Image Reconstruction Algorithms}\label{algo}
{\it RHESSI} is a Fourier imager employing nine rotating modulation collimators
to modulate the incident X-ray (and $\gamma$-ray) fluxes from the Sun,
which are recorded in nine electrically segmented
germanium detectors behind the collimators \citep{Lin02, Hurford02, Smith02}.
HXR images can be reconstructed with several algorithms of general purposes
\citep{Hurford02}, e.g., Clean, MEM\_NJIT, uv\_smooth, and Pixon.

The Clean algorithm assumes a flare image to be a superposition of point sources,
whose location and strength are called the Clean components,
and aims to remove the sidelobes of a back projection dirty map 
\citep{Hurford02, Dennis09}.
It iteratively selects the brightest point from the (dirty) residual map and 
subtract from the map a fraction of its flux multiplied with 
the point source function (PSF) to form a new residual map 
until the maximum flux becomes negative or the specified iteration number is reached.
In practice, the software convolves these Clean components with 
the Gaussian-shaped PSF (aka ``Clean beam") with 
addition of the final residual map to display the Clean image. 
So far Clean is the most commonly used method in the literature for {\it RHESSI} flares;
but sometimes the displayed images are rather diffuse.
The Pixon method aims to construct the simplest image consistent with the data 
and is regarded as the most photometrically accurate \citep{Metcalf96},
although it is much more time-consuming than other methods \citep{Aschwanden04}.
The MEM\_NJIT and uv\_smooth are algorithms based on the concept of visibility 
\citep{Hurford02}, which is the Fourier transforms of the source images,
instead of the time-binned modulation profiles for Clean and Pixon.
The MEM\_NJIT algorithm \citep{Schmahl07, Bong06} is a fast maximum entropy method 
(MEM) to maximize the information entropy 
and the MEM methods usually produce very sharp images,
which can help resolve sources close to each other \citep[e.g.][]{Chen05}.
The newly invented uv\_smooth method \citep{Massone09} interpolates a finite set of 
sparsely sampled visibilities and generate the images through Fourier inversion.
More details of these algorithms and comparison among them can be found in 
\citet{Hurford02}, \citet{Aschwanden04}, \citet{Dennis09}, and \citet{Massone09}.

\begin{figure*}
\epsscale{1.15}\plotone{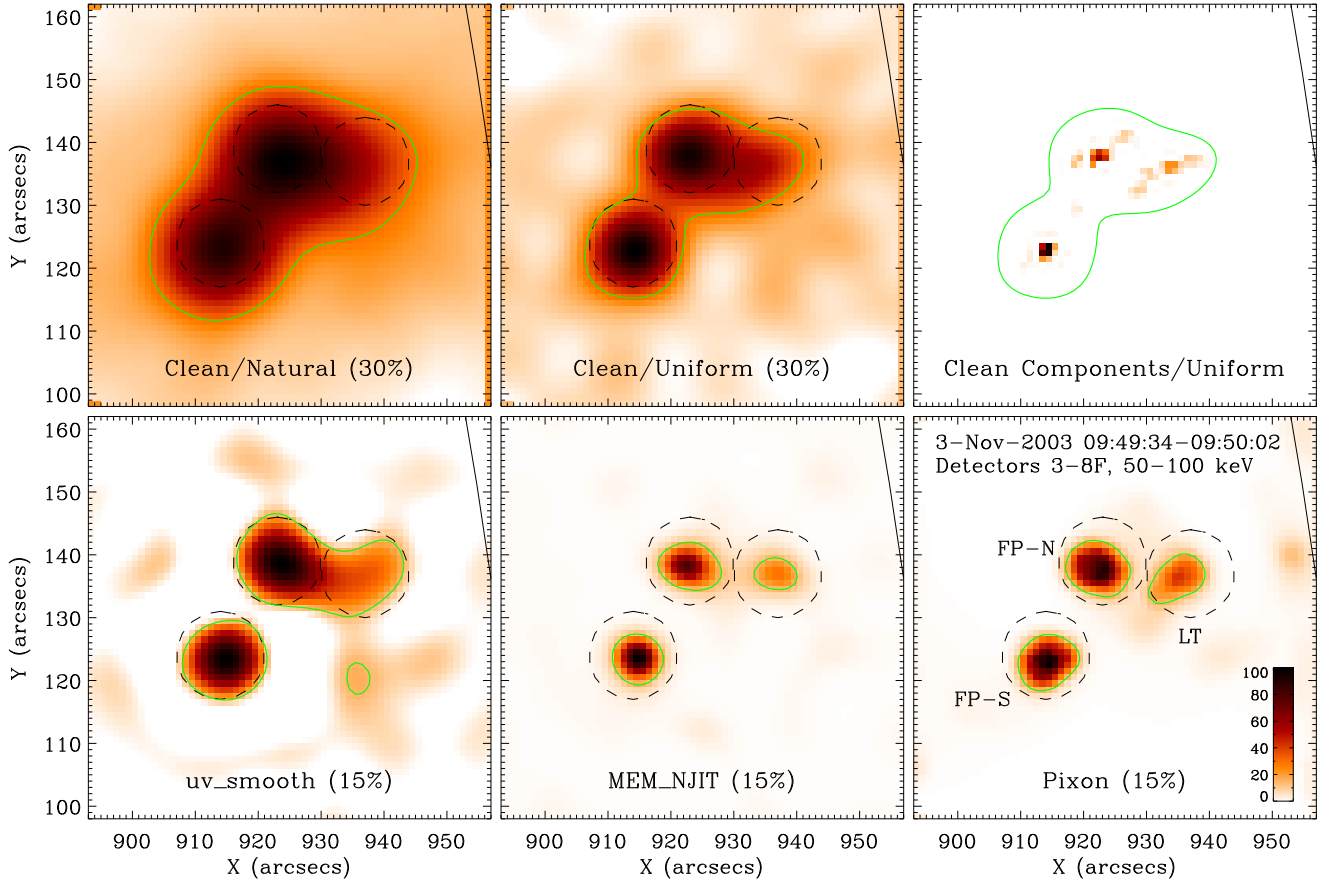}
\caption{HXR images at 50--100 keV 
during the nonthermal peak as reconstructed by different algorithms
from the front segments 3--8.
The contours superposed on the images are calculated relative to
the maximum intensity of each image.
The Clean images show two FP sources as previously found, 
while the MEM\_NJIT and Pixon images, and the Clean components as well,
clearly resolve an additional bright LT source.} 
\label{fig_algo}
\end{figure*}

The 2003 November 3 flare was earlier analyzed with the Clean method
\citep[see][]{LiuW04, Veronig06, Ishikawa11a}.
We reanalyze this flare by using all the above methods.
In Figure \ref{fig_algo} we present the HXR images at 50--100 keV
from the front segments 3--8 of FWHM $\sim$6.8\arcsec\ during the nonthermal peak. 
It is surprising that the MEM\_NJIT and Pixon images show not only two FP sources
but also a distinct LT source. 
Note that actually the Clean components also indicate a rather strong source 
at the LT location 
as seen in the Pixon and MEM\_NJIT images.
On the contrary, in the Clean images, 
the LT source appears to be part of an apparently 
elongated source, which was treated as the northern FP.
This is due to convolution of the Clean components with the PSF from the segments 3--8,
whose default beam width in the software may be too broad to resolve the LT source.
The uv\_smooth image shows slightly more prominent LT emission than the Clean images,
although still much less distinctive than the Pixon and MEM\_NJIT images.

We simply ascribe the finding of the high energy LT source 
to the ``super-resolution" power of 
the Pixon and MEM\_NJIT algorithms in resolving close sources.
As expected, inclusion of the first segment 
with the finest angular resolution of FWHM $\sim$2.3\arcsec\ does reveal the LT source 
in the Clean images (not shown here).

\section{Examination of Pulse Pileup Effect}\label{pileup}
Throughout the 2003 November 3 flare, both the thin and thick 
attenuators (the A3 state) are inserted in place at front of the detectors
to reduce the intense thermal emission.
The relatively low fractional livetime, $\sim$63\% averaged over the flare nonthermal peak
\citep[see also][]{Ishikawa11a},
indicates potential pulse pileup effect on this X3.9 class flare. 
A preliminary examination (based on the program hsi\_pileup\_check.pro in SSW)
indicates that $\sim$20--30\% of the spatially integrated
count rates above $\sim$50 keV are due to pileup. 
However, We note that 
this percentage only exceeds the LT contour level at 93--153 keV and is
lower than the LT contour levels below 93 keV as shown in Figure \ref{fig_pixon6}.

Pileup occurs when two or more (but with a much lower probability)
photons arrive at a detector almost simultaneously such that
they are registered indistinguishably as one single photon
whose energy is the sum of the individual photon energies \citep{Smith02, Hurford02}.
The probability of pileup is roughly proportional to the square of the count rate.
Its main effect is to produce spectral distortion, 
most significantly at energies twice the peak energy of the count rate spectrum
($\sim$18 keV in the A3 state).
For very high count rates, it can produce an artificial (``ghost") image
at high energy siting atop the main source of the peak energy. 
We note that, 
although there exist procedures for preliminary pileup correction of
the spatially integrated spectra \citep{Smith02} and for forward-modeling 
simulation in imaging spectroscopy \citep{Schwartz08, LiuW09a},
so far the pileup effect and its correction still remain a challenging topic
when analyzing imaging spectra.

We show here that the pileup effect 
should not be important on the distinct, high energy LT coronal source
as seen in the current flare. 
Our confidence relies on the appearance of the high energy LT source itself:
its unusually high energies and its large separation from the most intense part of 
the thermal loop.
First, pileup mostly comes from two thermal photons near the peak energy of
the count rate spectrum (in first order), therefore the ``ghost" image should appear 
at the same place as the thermal emission (R. A. Schwartz, private communication).
Nevertheless, we do not find such a source at energies $\ge$25--50 keV. 
Similarly, \citet{Saint-Hilaire08} claimed that the pileup effect has 
negligible influence on the FP spectra for those flares in which the FPs are 
spatially distinct from the thermal loop.
Second, the LT's highest energy is at least five times greater 
than the peak energy of the count spectrum, which makes it very unlikely for 
the pileup effect to generate a {compact} source at such high energies.
Note that the pileup effect tends to produce a rather diffuse 
``ghost" source (G. J. Hurford, private communication).
Although pileup in higher orders is more complicated and 
cannot be completely ignored, its effect on imaging 
has not been studied so far (S. Krucker, private communication).
Third, we note that the upper coronal source that is located away from the
thermal loop cannot be due to pileup because of its low thermal intensity. 
Since bremsstrahlung emissivity is mainly proportional to the local plasma density,
in this sense, 
it is reasonable to expect a more intense LT source at lower altitudes,
which is the focus of our study here.

Therefore we believe that the high energy LT source up to $\sim$100--150 keV
in the 2003 November 3 flare 
is physically real and not an artifact of the pileup effect.
We cannot find the LT source at higher energies ($\ge$150 keV)
based on the images reconstructed from the rear segments,
which are nearly unaffected by the pileup effect
\citep[see also][]{Ishikawa11a}.

\end{document}